\documentclass[a4paper,10pt,pra,twocolumn,showpacs,preprintnumbers,amsmath,amssymb]{revtex4}
\usepackage{amsmath}
\usepackage{amssymb}
\usepackage{graphicx}
\usepackage{dcolumn}
\usepackage{bm}
\usepackage[below]{placeins}

\begin{document}
\title{Pomeranchuk effect and spin-gradient cooling of Bose-Bose mixtures \\ in an optical lattice}
\author{Yongqiang Li$^{1,2}$, M. Reza Bakhtiari$^{1}$, Liang He$^{1}$, and Walter Hofstetter$^{1}$}
\address{$^{1}$Institut f\"ur Theoretische Physik, Johann Wolfgang
Goethe-Universit\"at, 60438 Frankfurt am Main, Germany \\
$^{2}$Department of Physics, National University of Defense Technology, Changsha 410073, P. R. China}


\begin{abstract}
We theoretically investigate finite-temperature thermodynamics and
demagnetization cooling of two-component Bose-Bose mixtures in a
cubic optical lattice, by using bosonic dynamical mean field
theory (BDMFT). We calculate the finite-temperature phase diagram,
and remarkably find that the system can be {\it heated} from the
superfluid into the Mott insulator at low temperature, analogous
to the Pomeranchuk effect in $^3$He. This provides a promising
many-body cooling technique.
We examine the entropy distribution in the trapped system and
discuss its dependence on temperature and an applied magnetic
field gradient. Our numerical simulations quantitatively validate
the spin-gradient demagnetization cooling scheme proposed in
recent experiments.
\end{abstract}

\date{\today}
\maketitle



\section{introduction}
Exploring the thermodynamics of interacting many-body systems has
been arguably one of the most important achievements of
cold-atomic gases, whether solely trapped by an external potential
or loaded into an optical lattice. A key experimental requirement
is reliable thermometry, with a precision below the degeneracy
point where quantum effects start to dominate. To date, the
temperature of a dilute (bosonic) gas in free space has been
measured by conventional time-of-flight thermometry \cite{D. C.
Mckay_2010} based on absorption imaging of an expanding gas
released from the trap. For a wide range of experiments, this
thermometric approach has been successful. However it loses its
applicability when the bosons are loaded into an optical lattice,
due to the reduced kinetic energy of the atoms \cite{D. C.
Mckay_2010,D. M. Weld}.
In a frequently used approximate approach, the temperature is
first measured in the absence of the optical lattice, which is
then ramped up gradually. One thus determines the final
temperature of the gas under the assumption of ramping up the
lattice adiabatically, which is a challenging task itself.
Therefore the need for an alternative approach is inevitable and
the search for other thermometers, directly applicable in optical
lattices, has been the subject of several theoretical proposals
\cite {D. C. Mckay_2010}.

One of the ultimate goals of experiments on cold-atomic gases in
optical lattices is to include the spin degree of freedom and to
simulate solid-state phenomena such as high-temperature
superconductivity whose underlying mechanism is still elusive
\cite{H_Tc_2002, Esslinger2010}. Recently, bosons with a spin
degree of freedom have been loaded into optical lattices \cite{J.
Catani_2008,D. M. Weld, B. Gadway} and significant efforts have
been made to achieve a magnetic phase transition in the
two-component bosonic system which has a rich phase diagram
\cite{Duan2003,Altman2003,Capogrosso2009,Hubener}. However, at
present it is still challenging to observe these quantum magnetic
phases in an optical lattice due to the extremely low critical
temperature which is governed by second-order tunneling \cite{A.
Auerbach, S. Folling, S. Trotzky, P. Medley, RBDMFT}. Different
cooling schemes have been proposed for lowering the temperature,
such as cooling based on extracting entropy from the region or
species of interests \cite{M. Popp, T.L.Ho_2009, T.L.Ho_2010, J.
Catani_2009, D. C. Mckay_2010}. Recently, a cooling approach using
spin-gradient adiabatic demagnetization was proposed in Ketterle's
group \cite{D. M. Weld}, and based on it a temperature of $350$
picokelvin has been achieved for a two-component Mott insulator of
$^{87}$Rb in a three-dimensional (3D) lattice \cite{P. Medley, D.
M. Weld_2010}. However, this temperature is still higher than the
critical temperature of the magnetic phase transition
\cite{entropy,RBDMFT}. In addition, direct evidence for the
validity of the spin-gradient cooling scheme is still lacking due
to severe approximations in the theoretical discussion \cite{D. C.
Mckay_2010}. As far as we know, the only theoretical simulation
related to the cooling of a two-component bosonic lattice gas in
the presence of a magnetic field gradient has been performed by
studying the domain wall dynamics of the Mott insulator via
mapping it onto a spin model, where a cooling effect is also
observed during adiabatic demagnetization \cite{Mueller}.

Due to isolation of the system from the environment, entropy is
more suitable than temperature for characterizing thermodynamical
properties. Entropy controls quantum phase transitions since it is
related to the number of accessible quantum states \cite {D. C.
Mckay_2010}. Therefore, a crucial issue related to cooling is how
entropy is distributed in the strongly interacting many-body
systems in current experiments \cite{J. Catani_2009, D. M. Weld,
P. Medley, B. Gadway} and how the entropy redistributes during the
adiabatic process of spin-gradient cooling \cite{P. Medley}. These
questions motivated our study in this paper focusing on the
thermodynamical properties of two-component Bose gases in optical
lattices in the presence of an external harmonic trap. While the
thermodynamics of strongly interacting two-component Fermi gases
has been investigated in detail \cite{Hulet, Jin, Luo, Bulgac,
Nascimbene1, Nascimbene2, Horikoshi, Hu, Esslinger,F. Werner, T.
Paiva} and the resulting critical entropy per particle $s \approx
k_B \ln 2$ at the fermionic Mott-insulator transition has been
achieved experimentally in a 3D cubic lattice
\cite{Bloch_Mott_phase, Esslinger}, less attention has been paid
to the thermodynamics of two-component bosonic systems \cite{Guertler}. In
Ref.~\cite{entropy}, the critical entropy for magnetic ordering of
two-component $hard$-$core$ bosons has been investigated in a 3D
homogeneous system, where a critical entropy per particle of
$0.35k_B$ for the XY-ferromagnetic phase and $0.5k_B$ for the
Z-N\'eel antiferromagnetic phase have been found. Here, we will
focus on the thermodynamical properties of realistic two-component
bosons in a 3D cubic optical lattice \emph {in the presence of an
external trap}, and investigate the validity of spin-gradient
demagnetization cooling, which is in principle capable of cooling
the system down to the critical temperature of magnetic order.
This system can be approximately described by a single-band
Bose-Hubbard model and is investigated by bosonic dynamical mean
field theory (BDMFT) \cite{Vollhardt, Hubener, Tong, Werner}, both
in combination with a local density approximation (LDA) and by its
full real-space implementation \cite{RBDMFT}.

The paper is organized as follows: in section II we give a
detailed description of the model and our approach used to
calculate the entropy in the inhomogeneous system. In section III
we present the finite temperature phase diagram of a Bose-Bose
mixture in an optical lattice. We then discuss the entropy
distribution in the presence of a harmonic trap without magnetic
field gradient. Finally we give a detailed discussion of the case
where a magnetic field gradient is applied. We conclude in section
IV.

\section{Model and Method} \label{sec:model}
We consider two species of bosonic atoms \cite {J. Catani_2008}
or, alternatively, atoms in two different hyperfine states
\cite{D. M. Weld, B. Gadway}, in an optical lattice in the
presence of an external harmonic trap. Within the tight-binding
picture, this system can be described by a single-band
Bose-Hubbard model:
\begin{eqnarray} \label{Hamil}
 \mathcal{H}=&-& \sum_{\stackrel{<i,j>}{\nu=b,d}} t_\nu (b^\dagger_{i\nu}b_{j\nu}+h.c)+\frac{1}{2}\sum_{i,\lambda\nu} U_{\lambda\nu} \hat{n}_{i\lambda}
(\hat{n}_{i\nu}-\delta_{\lambda\nu}\nonumber) \\
&+&\sum_{i,\nu=b,d} (V_i-\mu_\nu)\hat{n}_{i\nu} - \sum_{i,\nu}
\mu^\nu_{mag}B(x_i)\hat{n}_{i\nu}
\end{eqnarray}
Moreover, we consider a linear position-dependence of the magnetic
field in $x$ direction, {\it i.e.}, $B(x_i)=c\, x_i$ where $c$ is
the magnetic field gradient and $x_i$ the distance from the
harmonic trap center, which describes the recent experiment \cite
{P. Medley}. This leads to
\begin{eqnarray}
 H_B=-\sum_{i,\nu} \mu^\nu_{mag}\, B(x_i)\hat{n}_{i\nu}=-\sum_{i,\nu} \mu^\nu_{mag} c\, x_i\hat{n}_{i\nu}\nonumber \\
 \equiv -\sum_{i,\nu} V^\nu_{grad} x_i\hat{n}_{i\nu}
\end{eqnarray}
In the Hamiltonian, $\langle i,j\rangle$ denotes the summation
over nearest neighbors sites and the two boson species are
labelled by the index $\lambda (\nu)=b,d$. Due to different masses
or a spin-dependent optical lattice, these two species generally
have different hopping amplitudes $t_b$ and $t_d$. The bosonic
creation (annihilation) operator for species $\nu$ at site $i$ is
$b^{\dagger}_{i\nu}$ ($b_{i\nu}$) and the local density is
$\hat{n}_{i,\nu}=b^\dagger_{i\nu}b_{i\nu}$. $U_{\lambda\nu}$
denotes the inter- and intra-species interactions, which can be
tuned via a Feshbach resonance \cite{A. Widera} or by a
spin-dependent lattice \cite{B. Gadway}. $\mu_\nu$ denotes the
global chemical potential for the two bosonic species and $V_i$ is
the harmonic trapping potential. $\mu^\nu_{mag}$ denotes the
magnetic moment of component $\nu$ and $B(x_i)$ is the magnetic
field along the $x$ axis.

Bosonic DMFT (BDMFT) has been developed \cite{Vollhardt} and
implemented \cite{Hubener, Tong, Werner} to provide a
non-perturbative description of zero- and finite-temperature
properties of the homogeneous Bose-Hubbard model including
magnetic ordering. In order to account for the external trapping
potential, we have recently developed real-space BDMFT (RBDMFT),
whose detailed formalism is presented in \cite{RBDMFT}. In
parallel to RBDMFT, here we also employ an LDA scheme combined
with single-site BDMFT to explore the system. The advantage of the
latter approach is the larger system size accessible. The validity
and limitations of this approach have been investigated by a
quantitative comparison with the more rigorous RBDMFT method
\cite{RBDMFT}. In our LDA+BDMFT calculations, the chemical
potentials are adjusted locally according to the trapping
potential, i.e., $\mu_\nu(r)=\mu_\nu-V_0r^2$, where $V_0$ is the
strength of the harmonic confinement and $r$ is the distance from
the trap center.

In general, it is difficult to calculate the entropy within BDMFT
or RBDMFT directly. But assuming that the strongly interacting
many-body system is in equilibrium, we can use the Maxwell
relation $\frac{\partial s}{\partial \mu} = \frac {\partial
n}{\partial T}$ to obtain the local
entropy per site~\cite{Bloch_Mott_phase} at temperature $T$ and chemical
potential $\mu_s(r)=(\mu_b(r)+\mu_d(r))/2$:
\begin{equation}\label{entropy}
s(\mu_s(r_0),T)=\int_{-\infty}^{\mu_s(r_0)} \frac {\partial n
(r)}{\partial T}d\mu_s (r)
\end{equation}
where $n (r)=n_b+n_d$ is the local density ({\it i.e.}, number of
particles per lattice site) at radius $r$. Note that the formula
(\ref{entropy}) is only valid at fixed $\Delta \mu(r)=
\mu_b(r)-\mu_d(r)$ for the two-component mixture. The density
distribution obtained from BDMFT and RBDMFT is accurate enough to
yield precise results for the derivative $\frac {\partial n
}{\partial T}$. This relation will be used in the following to
obtain the entropy distribution.

\FloatBarrier
\section{Results}
In ongoing experiments, two hyperfine states of $^{87}$Rb have
been loaded into optical lattices \cite{D. M. Weld, B. Gadway},
with inter-species and intra-species interactions in the regime
$U_b \approx U_d \approx U_{bd}$. Considering the tunability of
interactions via Feshbach resonances \cite{A. Widera} or
state-dependent optical lattices \cite{B. Gadway}, here we choose
$U_b=U_d = 1.01U_{bd}$. In the following, we investigate the
finite temperature quantum phases of this system in a cubic
optical lattice, as well as the temperature dependence of the
entropy distribution in the presence of a harmonic trap. Finally,
these thermodynamical properties are used to quantitatively
describe the adiabatic spin gradient cooling scheme of \cite{P.
Medley}. In all our calculations we consider balanced mixtures of the two components. We choose $U_{bd} = 1$ as the
unit of energy, and set $k_B=1$. $z$ denotes the number of nearest
neighbors for each lattice site. The lattice constant is set to
unity.

\subsection{Pomeranchuk effect and phase diagram at finite temperature}
\begin{figure}[h]
\vspace{-20pt} \hspace{-0pt}
\begin{tabular}{ c }
\includegraphics*[width=4.5in] {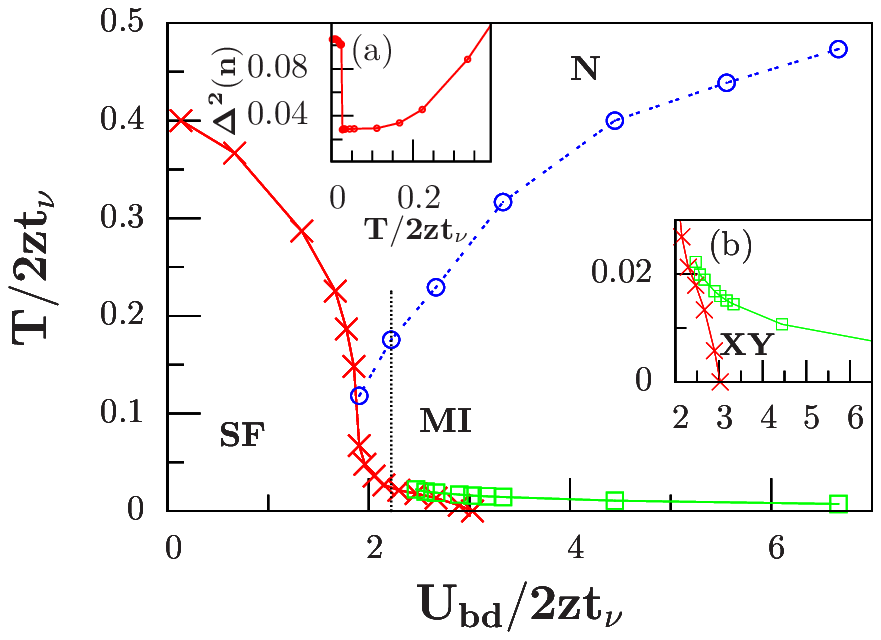}
\vspace{-40pt}
\\
\includegraphics*[width=4.5in] {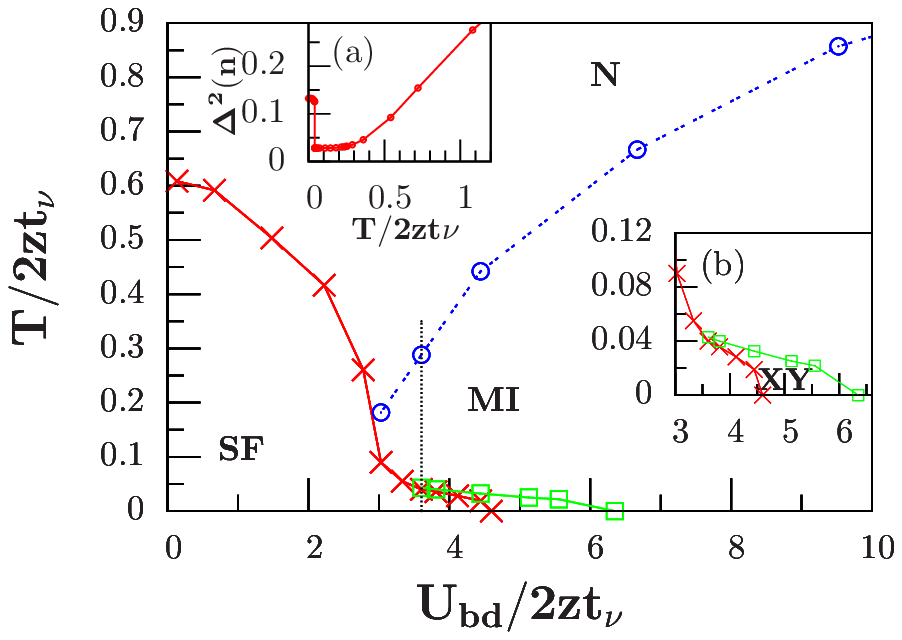}
\end{tabular}
\vspace{-33pt} \caption{(Color online) Finite temperature phase diagram of a
two-component bosonic gas in a cubic optical lattice with filling
$n_b=n_d=0.5$ ({\bf upper}) and $n_b=n_d=1.0$ ({\bf lower}). The
interactions are set to $U_b=U_d=1.01U_{bd}$, and the hopping
amplitudes are $t_b=t_d$. Inset (a): fluctuations of the total
density per site $n=n_b+n_d$ as a function of temperature along the
vertical dotted line of the main figure. Note the reduction of
local number fluctuations by heating, analogous to the \emph
{Pomeranchuk effect}. Inset (b): zoom of the main figure around
the critical point of magnetic order.}\label{phase}
\end{figure}
In this section, we explore the finite-temperature phase diagram of two-component bosons in a homogeneous and infinite optical lattice. For strongly interacting two-component Fermi gases, the critical
parameters such as the critical temperature and entropy for the
transition to a superfluid state have been determined
experimentally \cite{Luo} by considering entropy versus energy.
For one-component bosonic gases in an optical lattice, the
finite-temperature phase diagram has been studied experimentally
in combination with Monte Carlo simulations \cite{S.
Trotzky_2009}. However, for two-component bosonic gases, the
critical behavior of the superfluid-normal phase transition has
not been determined yet. In our previous work \cite{RBDMFT}, phase
diagrams at filling $n=1$ and $n=2$ for zero and fixed finite
temperature have been determined for the cubic lattice. But there
we mainly focused on the emergence of long-range magnetic order,
which is governed by second-order tunneling and only develops at
very low temperatures of the order of $100$ pK. On the contrary,
here we will investigate quantum criticality of the system at
higher temperatures. We choose interactions $U_b=U_d=1.01U_{bd}$
and hopping amplitudes $t_b=t_d$. Fig.~\ref{phase} shows the phase
diagram of a Bose-Bose mixture in a cubic optical lattice with
filling $n_b=n_d=0.5$ (upper) and $n_b=n_d=1$ (lower). We observe
four different phases. When the interaction is weak, the atoms are
delocalized and at low temperature the system is in the superfluid
phase (SF), characterized by a finite value of the superfluid
order parameter $\phi_\nu\equiv\langle b_\nu \rangle$. When the
temperature is increased, thermal fluctuations destroy the
coherence between atoms and the system goes through a phase
transition into the normal phase (N). For sufficiently strong
interactions, the atoms are localized and hopping processes are
strongly suppressed. The system is in the XY-ferromagnetic phase
(characterized by $\langle bd^\dagger\rangle > 0$ and
$\phi_\nu=\langle b_\nu \rangle = 0$) at low temperature, with
magnetic long-range order governed by second-order tunneling
processes. Since the corresponding energy scale is very small,
even weak thermal fluctuations can destroy the long-range magnetic
order, and the system will go through a phase transition into a
Mott insulator (MI) without order. Upon further increase of
temperature, the Mott insulator melts into a normal phase which is
characterized by large density fluctuations $\Delta^2 (n)=\langle
(n - \langle n \rangle)^2 \rangle$ where the $n$ is the total
density per site. Compared to the single-component system in a
cubic optical lattice, new features of two-component bosons appear
at low temperature. Near the critical interaction strength of the
zero-temperature MI-SF transition, with increasing temperature,
the system will first go through a phase transition from
superfluid to Mott-insulator, and then cross over to the normal
phase. This is because upon heating at low temperature, the system
favors localization - analogous to the Pomeranchuk effect in
liquid $^3$He \cite{R. C. Richardson, F. Werner} - since the Mott
insulating phase of spinful bosons carries more entropy in the
spin degree of freedom than the superfluid. Interestingly, the
first-order phase transition from superfluid to Mott-insulator
occurs at a higher temperature for filling $n = 2$ (lower plot in
Fig. \ref{phase}) compared to $n = 1$, indicating that it is
easier to observe the Pomeranchuk effect discussed above for
higher filling. Note that the XY-ferromagnetic phase at filling
$n=2$ only extends up to a finite maximum value of
$U_{bd}/2zt_\nu$, which is consistent with our previous work
\cite{RBDMFT}.

\subsection{Entropy distribution in the trapped system with $B=0$}

In the former section, we have studied the homogeneous system and
mapped out the finite-temperature phase diagram. We will now study
the thermodynamics of Bose-Bose mixtures in an optical lattice in
the presence of a harmonic trap. More specifically, we investigate
the temperature dependence of the entropy distribution, motivated
by recent experiments  \cite{J. Catani_2008, D. M. Weld, B.
Gadway}. Comparison between RBMDFT and BDMFT+LDA calculations has
been made to check the validity of  LDA for determining the
entropy. Only the results of BDMFT+LDA are given here for the 3D
case in a $51 \times 51 \times 51$ cubic lattice. Throughout this
section, the interactions are set to $U_b=U_d=1.01U_{bd}$ with a
harmonic trap strength $V_0=0.005U_{bd}$ and a total filling $n=2$
at the trap center.
\begin{figure}[h]
\begin{tabular}{ c }
\includegraphics*[width=3.2in]{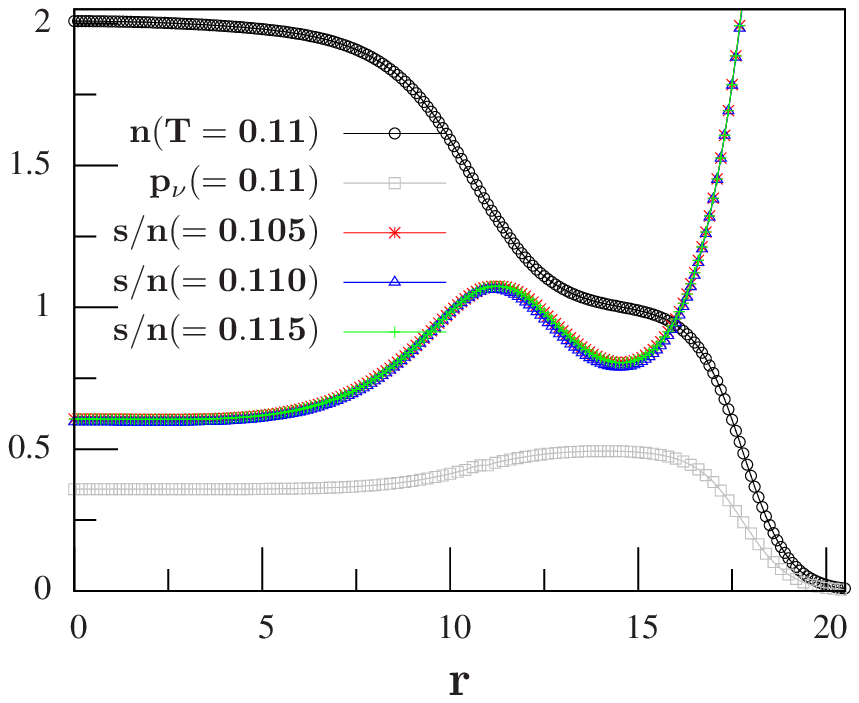}
\\
\includegraphics*[width=3.2in]{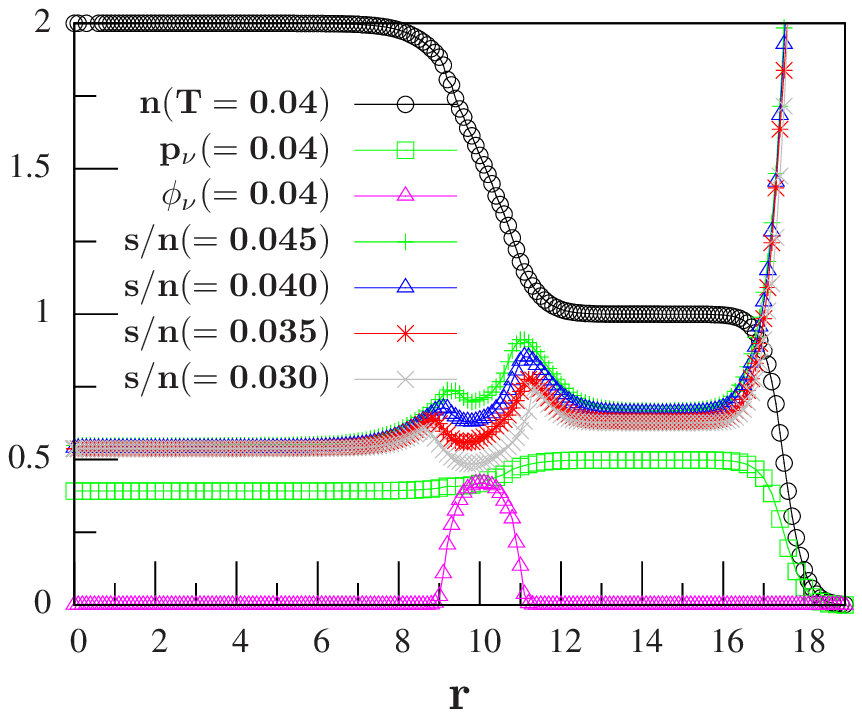}
\end{tabular}
\caption{(Color online) Radial profile for the total density per site ($n \equiv n_b+n_d$), parity ($p_\nu$), local entropy per particle ($s/n$), and
superfluid order parameter ($\phi_\nu$) in a 3D cubic
lattice obtained by BDMFT+LDA at different temperatures. The
interactions are set to $U_b=U_d=1.01U_{bd}$, with hopping
amplitudes $2zt_b=2zt_d=0.195U_{bd}$ and harmonic trap strength
$V_0=0.005U_{bd}$. The unit of temperature is $U_{bd}$.}
\label{3D_entropy}
\end{figure}

In the upper panel of Fig.~\ref{3D_entropy} at high temperatures
$T/U_{bd}=0.105$, $0.11$ and $0.115$ (corresponding to the normal
phase in Fig.~\ref{phase}), the Mott-insulator plateaux melt into
a normal phase with entropy per site $s>\ln 3$ around the center
of the harmonic trap and $s>\ln 2$ at the second Mott-insulating
ring. Naturally, we can also identify the melting of the Mott
insulator into the normal phase from the density profile, i.e.,
the corresponding Mott-plateaux at filling $n=2$ and $n=1$ have
disappeared at this temperature. Due to the insensitivity of the
density profile to a small variation of temperature, only a single
density profile at temperature $T/U_{bd}=0.11$ is shown here.
There are also two peaks of the entropy density in the normal
shells surrounding the Mott-insulating regions. Our simulations
indicate that the transfer of entropy from superfluid to Mott
insulator due to the Pomeranchuk effect does not occur in this
high temperature region, since here the local entropy per particle in
the superfluid is higher than in the Mott-insulator. We observe
that the local entropy per particle is reduced when the temperature decreases,
as shown in the lower panel of Fig.~\ref{3D_entropy} at low
temperatures of $T/U_{bd}=0.035$, $0.04$ and $0.045$
(corresponding to the Mott insulator region in Fig.~\ref{phase}).
Here the system has a Mott-insulator core with filling $n=2$ in
the trap center and also a Mott-insulating shell with filling
$n=1$. Correspondingly, the local entropy per site of the Mott-insulator
region is $s\approx \ln 3$ in the filling $n=2$ region and
$s\approx \ln 2$ in the $n=1$ region, respectively, since there
are three possible local spin states $\left |\uparrow \uparrow
\right \rangle$, $\left|\downarrow \downarrow \right\rangle$ and
$\left|\uparrow \downarrow \right\rangle$ for $n=2$, and two
possible spin states $\left|\uparrow \right \rangle$,
$\left|\downarrow \right\rangle$ for $n=1$, where $\uparrow$ and
$\downarrow$ denote the two bosonic species. Between the two
Mott-insulating regions, there is also a superfluid shell with
non-zero value of the superfluid order parameter. Interestingly,
we observe a sudden drop of the entropy density around the peak of
the superfluid order parameter, which indicates a fine structure
in the density distribution of the phases with non-integer filling
(superfluid and normal phase). A similar structure is also found
for a one-component Bose gas in an optical lattice plus external
harmonic trap \cite{Ho_2007}. Physically, the sudden change of
entropy in the superfluid region is caused by the reduced number
of many-body states of the system due to the formation of a
condensate. It is expected that, if the temperature is lowered
further, another superfluid domain forms in the region with
filling $n<1$. We have also shown the parity profile $p_\nu=\langle (1-e^{i\pi \hat{n}_\nu})/2\rangle$ for the individual components in Fig.~\ref{3D_entropy}, which can be directly measured experimentally~\cite{M. Greiner_2009, I. Bloch_2010}. Interestingly, the local parity for the individual components in the Mott-insulating region with total filling $n=2$ is finite.

In addition, we now observe (lower plot of Fig. \ref{3D_entropy})
that the local entropy per particle in the first superfluid ring is
smaller in some regions than that in the Mott insulator, which indicates that a transfer of entropy from
superfluid to Mott insulator can lower the temperature of the
system in this regime, which is consistent with the phase diagram
for the homogeneous system in Fig. \ref{phase}. This
interaction-induced cooling mechanism (Pomeranchuk effect) of
two-component bosonic gases in an optical lattice is expected to
be visible experimentally \cite{J. Catani_2008, D. M. Weld, B.
Gadway}, after further lowering the temperature. For example, in the experiment this effect could be observed via ramping up the optical lattice, where the temperature should be decreased beyond single-particle adiabatic cooling due to the Pomeranchuk effect, since the Mott-insulating region increases.

\begin{figure}[h]
\begin{tabular}{ c }
\includegraphics*[width=3.2in]{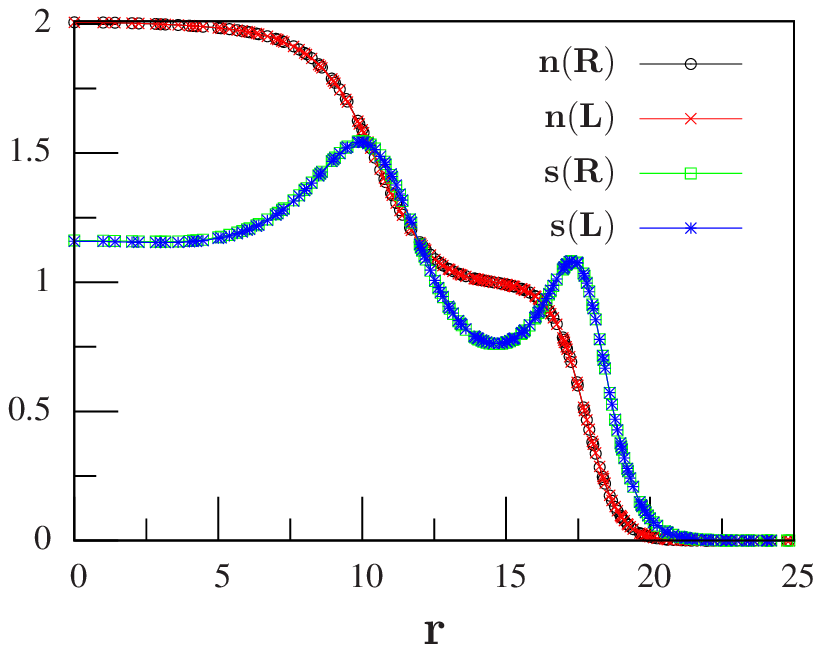}
\\
\includegraphics*[width=3.2in]{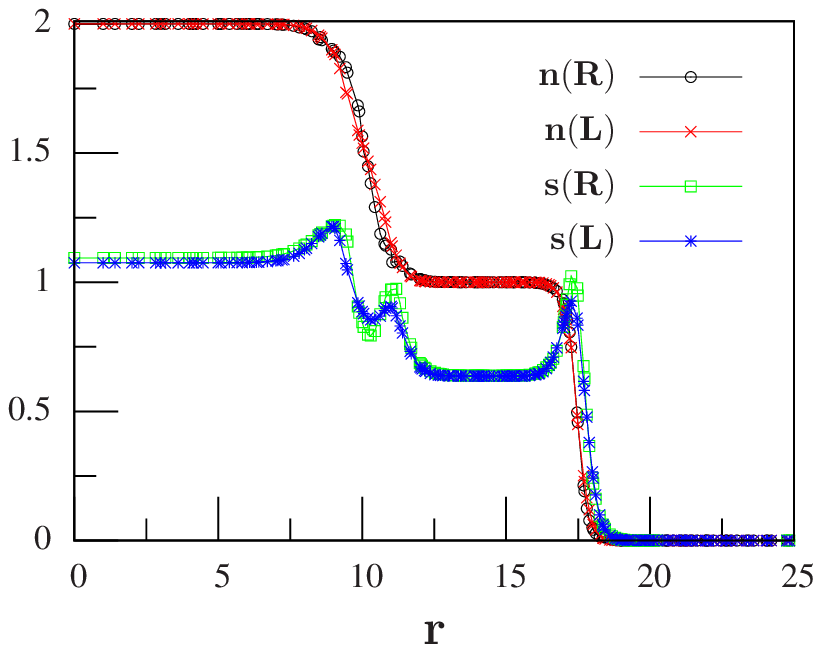}
\end{tabular}
 \caption{(Color online) Validity of BDMFT+LDA benchmarked against RBDMFT. Density profile $n_{tot}$ and entropy distribution $s$
 along the radial direction $r$ at temperature $T=0.1U_{bd}$ (\textbf {upper})
 and $T=0.03U_{bd}$ (\textbf {lower}) obtained by
RBDMFT (R) and BDMFT+LDA (L) for 2D case. The interactions are set
to $U_b=U_d=1.01U_{bd}$ and the hopping amplitudes are
$2zt_b=2zt_d=0.175U_{bd}$ with harmonic trap $V_0=0.005U_{bd}$.
}\label{2D_temperature}
\end{figure}
To check the validity of BDMFT+LDA around quantum degeneracy, we
investigate the density and entropy distribution for the 2D case
and test the accuracy of BDMFT+LDA against RBDMFT, as shown in
Fig.~\ref{2D_temperature}. We find excellent agreement deep inside
each phase, while RBDMFT provides the slightly more accurate
description of the transition region. We therefore expect that
BDMFT+LDA will also give quantitatively reliable results for the
3D case.

\FloatBarrier
\subsection{Adiabatic cooling via entropy redistribution for $B\neq 0$}

We have so far investigated thermodynamical properties for equal
filling of the two components. In this section, we will now study
a scenario with the two species separated by a magnetic field with
constant gradient which can be used experimentally to cool the
system. Specifically, we simulate the adiabatic process of the
spin-gradient cooling scheme proposed by Weld \emph{et.
al.}~\cite{D. M. Weld}. To this end, we calculate the entropy
distribution of the inhomogeneous system in the presence of the
field gradient, and the dependence of the entropy per particle on
temperature. To simplify the calculation, we assume that the two
components of the bosonic mixture have the same absolute value of
the magnetic moment.

\FloatBarrier

\subsubsection{Entropy distribution in the presence of field gradient}
\begin{figure}[h]
\includegraphics[width=3.2in]{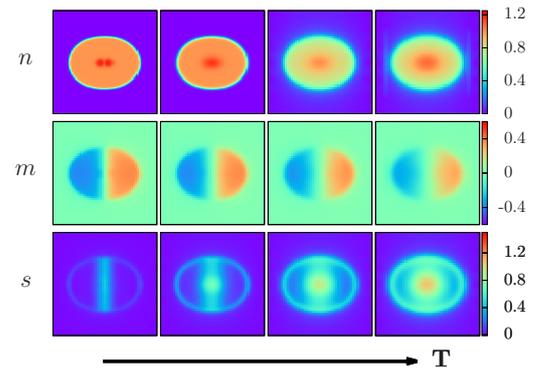}
 \caption{(Color online) Real-space profile (here in the $x$-$y$ plane of the lattice) for the local density $n$, magnetization $m$ and entropy $s$ along the $z=0$ plane of a 3D cubic lattice using BDMFT+LDA. From left to right, the temperatures are $T/U_{bd}=0.020$, $0.040$, $0.070$ and $0.095$, respectively. The interactions are set to $U_b=U_d=1.01U_{bd}$ and the hopping amplitudes are $2zt_b=2zt_d=0.12U_{bd}$, with total particle number $N_{tot}\approx 17000$ in a harmonic trap $V_0=0.004U_{bd}$ and magnetic field gradient $V_{gra}=0.01U_{bd}$.}\label{3D_entropy_extract_2D}
\end{figure}
The two-component bosonic mixture can be separated to opposite
sides of the trap by the magnetic field. At zero temperature, the
two components are completely separated and a sharp domain wall
forms in the trap center. At finite temperature, spin excitations,
such as a pair of opposite-spin atoms swapping positions via
second-order tunneling, will broaden the width of the domain wall (the width is defined as the distance from the trap center to the position where the magnetization is half of the maximum value).
As pointed out in \cite{D. M. Weld, Mueller} the width of the domain wall depends in a simple way on the field gradient and can be used as a thermometer in the zero-tunneling limit. Fig.~\ref{3D_entropy_extract_2D} shows the distribution of local density $n$, magnetization
$m=(n_b-n_d)/2$, and entropy $s$ in the $z=0$ plane. Since the
density and magnetization distributions depend on the temperature,
they can be used for thermometry via in-situ measurements with
single-site resolution \cite{M. Greiner_2009, I. Bloch_2010}.
In particular, the magnetization distribution can be used as a
thermometer at low temperatures down to the critical temperature
of magnetic phases. From the middle row of
Fig.~\ref{3D_entropy_extract_2D} we observe that the narrow mixed
region of the two components broadens with increasing temperature,
which is consistent with measurements where temperatures as low as
$350$ pK have been measured \cite{D. M. Weld,P. Medley}. The
bottom row of Fig.~\ref{3D_entropy_extract_2D} shows the entropy
distribution. The entropy is mainly carried by the spin degree of
freedom of particles around the trap center, and also by
delocalized particles near the edge of trap. When the temperature
is lowered, the delocalized particles form a condensate. As a
result, the entropy drops quickly as a function of temperature in
the superfluid ring. On the other hand, the spin degree of freedom
in the mixed region can still carry a large amount of entropy,
even at low temperature where the entropy of the single-component
superfluid becomes very small. Therefore, if one prepares the
system in a state where entropy is mainly carried by a single
species ({\it i.e.} if one initially separates the two species by
a field gradient) and then transfers the entropy from the single
species to the spin degree of freedom, the temperature of the
system can be lowered dramatically.

\begin{figure}[h]
\vspace{-3pt}
\begin{tabular}{ cc }
\includegraphics[scale=.51] {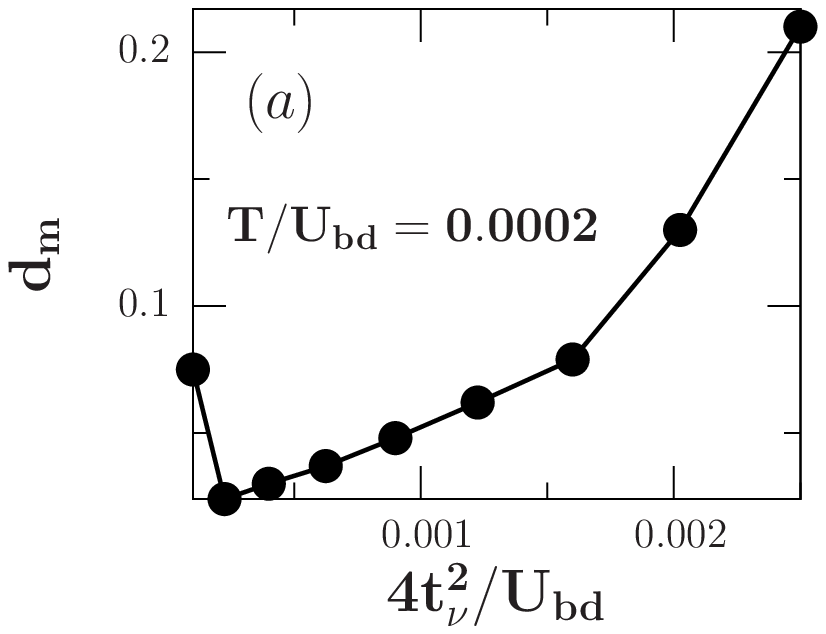}&
\hspace{-15pt}
\includegraphics[scale=.51]{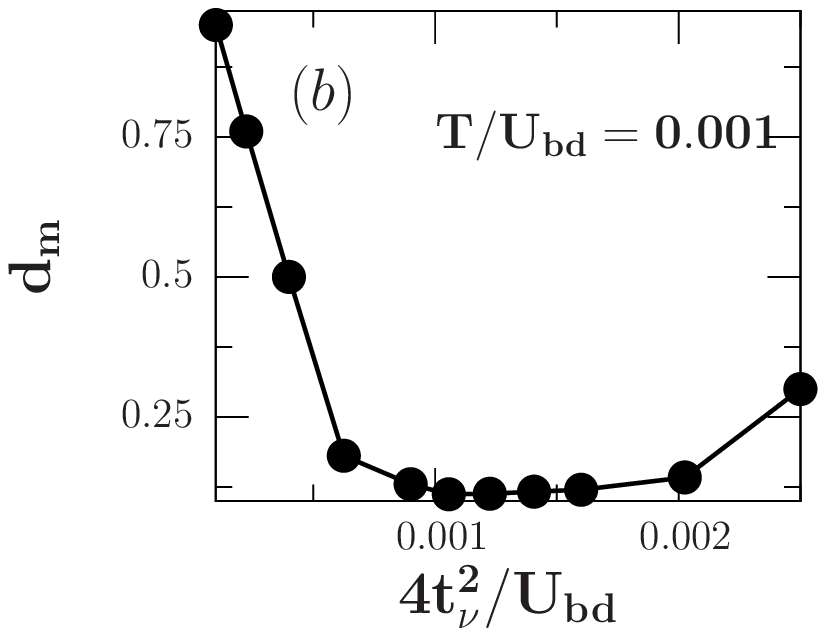}
\\
\includegraphics[scale=.51] {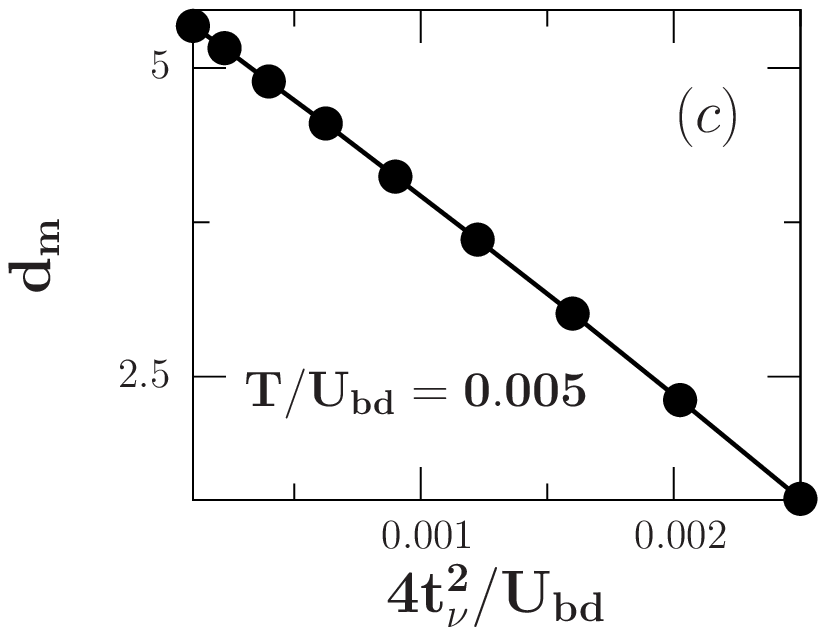}&
\hspace{-15pt}
\includegraphics[scale=.51]{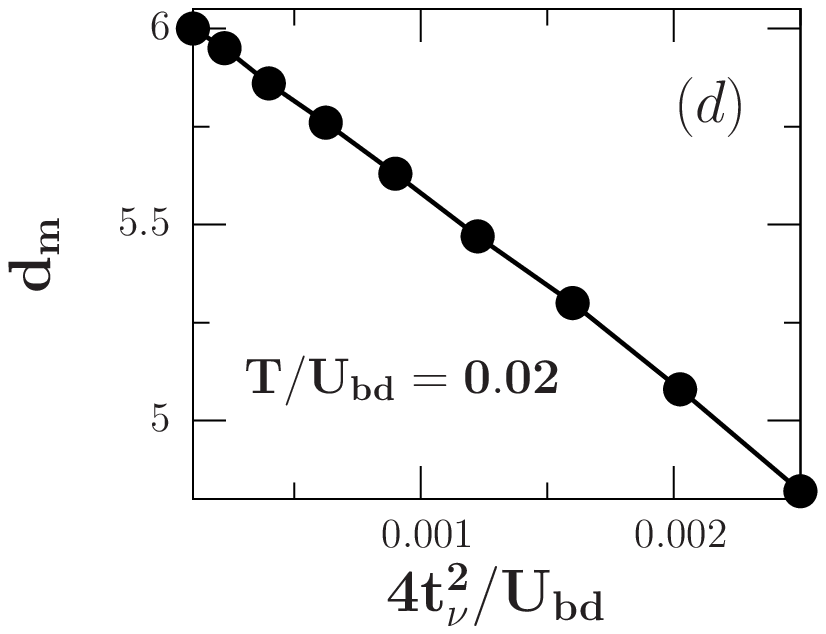}
\end{tabular}
\caption{Domain-wall width $d_m$ (in units of the lattice constant) as a function of superexchange coupling at different temperatures. The width is defined as the distance from the trap center to the position where the magnetization is half of the maximum value. The interactions are set to $U_b = U_d = 1.01U_{bd}$ in a harmonic trap $V_0 = 0.004U_{bd}$ and a magnetic field gradient $V_{gra} = 0.0005U_{bd}$.}
\label{width}
\end{figure}
\begin{figure}[h]
\includegraphics[width=3.in]{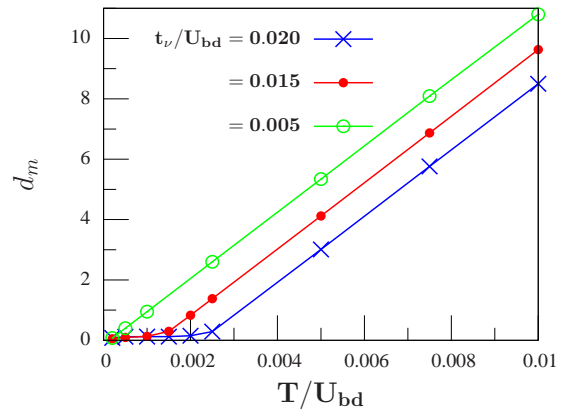}
\caption{(Color online) Domain-wall width $d_m$ (in units of the lattice constant) as a function of temperature at different hopping amplitudes. The width is defined as the distance from the trap center to the position where the magnetization is half of the maximum value. The interactions are set to $U_b = U_d = 1.01U_{bd}$ in a harmonic trap $V_0 = 0.004U_{bd}$ and a magnetic field gradient $V_{gra} = 0.0005U_{bd}$.}
\label{width_T}
\end{figure}

The domain-wall width can also be used as a tool to measure the strength of the resulting superexchange interactions between the atoms. As shown in panel (a) and (b) in Fig.~\ref{width}, when superexchange interactions dominate over thermal fluctuations ($4t^2_\nu/U_{bd}>T$), we observe a linear dependence of the domain-wall width on the strength of the superexchange in the Mott-insulating regime. We also observe that the domain-wall width increases faster at larger hopping parameters, since in that case the mixed region is in the superfluid regime and the first-order tunneling dominates. When thermal fluctuations dominate ($4t^2_\nu/U_{bd}<T$), as shown in panels (a), (b), (c) and (d), the increase of the superexchange decreases the width of the domain wall due to minimizing the energy of the spin-spin coupling. If the temperature is increased, the minimum of the domain-wall width is shifted to higher hopping amplitudes, as shown in panels (a) and (b) in Fig.~\ref{width}.
We also observe that the linear dependence~\cite{D. M. Weld, Mueller} of the domain-wall width $d_m$ only holds for temperature above the critical values $T_c$ for magnetic ordering (see Fig.~\ref{width_T}). The change of slope at $T_c$ is a clear indication of the phase transition.

\FloatBarrier
\subsubsection{Entropy per particle versus temperature}
\begin{figure}[h]
\includegraphics[width=3.2in]{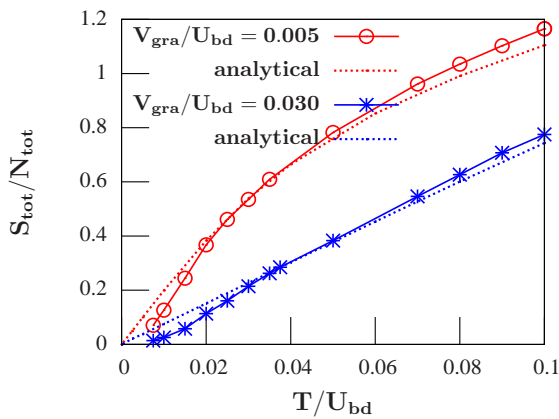}
\caption{(Color online) Entropy per particle versus temperature in a cubic
optical lattice obtained by BDMFT+LDA, compared with the
analytical zero-tunneling approximation~\cite{D. M. Weld_2010}.
The interactions are set to $U_b=U_d=1.01U_{bd}$ and the hopping
amplitudes to $2zt_b=2zt_d=0.12U_{bd}$, with total particle number
$N_{tot}\approx17000$ in a harmonic trap of strength
$V_0=0.004U_{bd}$. }\label{entropy_T}
\end{figure}
We now focus on the relation of entropy versus temperature, which
gives insight how adiabatic changes affect the temperature of the
system. Fig. \ref{entropy_T} shows the entropy-temperature curve
for strongly interacting two-component bosons in an optical
lattice in the presence of the magnetic field, where the dashed
lines are obtained by the zero-tunneling approximation~\cite{D. M.
Weld_2010}. Due to the deep optical lattice, our results obtained
by BDMFT+LDA are in good agreement with the approximate analytical
results except at low and high temperatures. At high temperature,
thermal fluctuations will induce hopping of atoms. This effect is
neglected in the zero-tunneling approximation, which therefore
gives a lower prediction for the entropy. At low temperature, on
the other hand, the entropy of the motional degree of freedom
drops quickly due to condensate formation in the superfluid
regime. This effect is neglected as well in the zero-tunneling
approximation, which therefore gives a larger prediction for the
entropy. We note that quantum Monte Carlo simulations
\cite{entropy} also reveal the inadequacy of the zero-tunneling
approximation in the low temperature regime.

\subsubsection{Adiabatic cooling via spin-gradient demagnetization}
\begin{figure}[h]
\begin{tabular}{ c }
\hspace{9.5pt}
\includegraphics[width=3.19in]{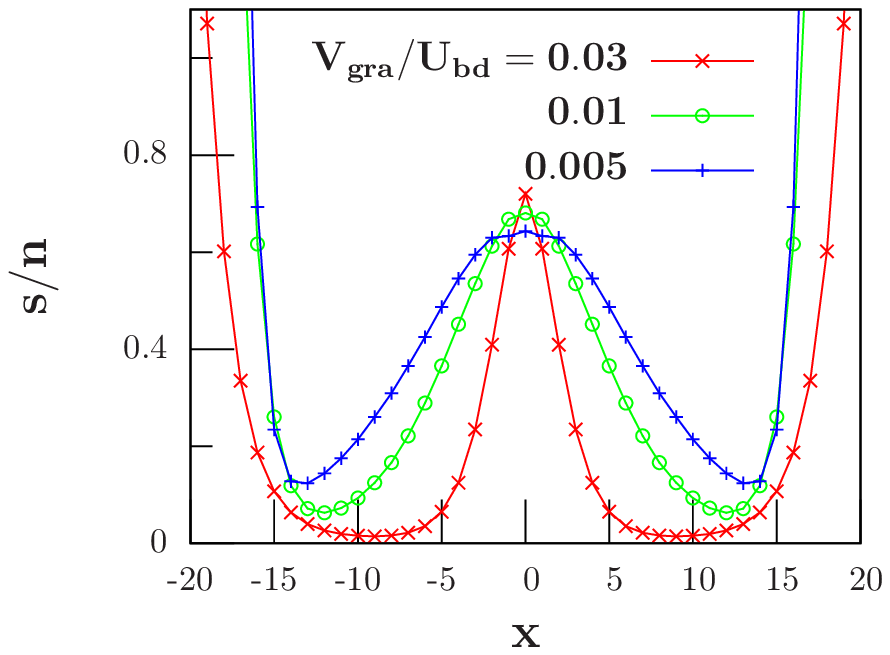}
\\
\includegraphics[width=3.4in]{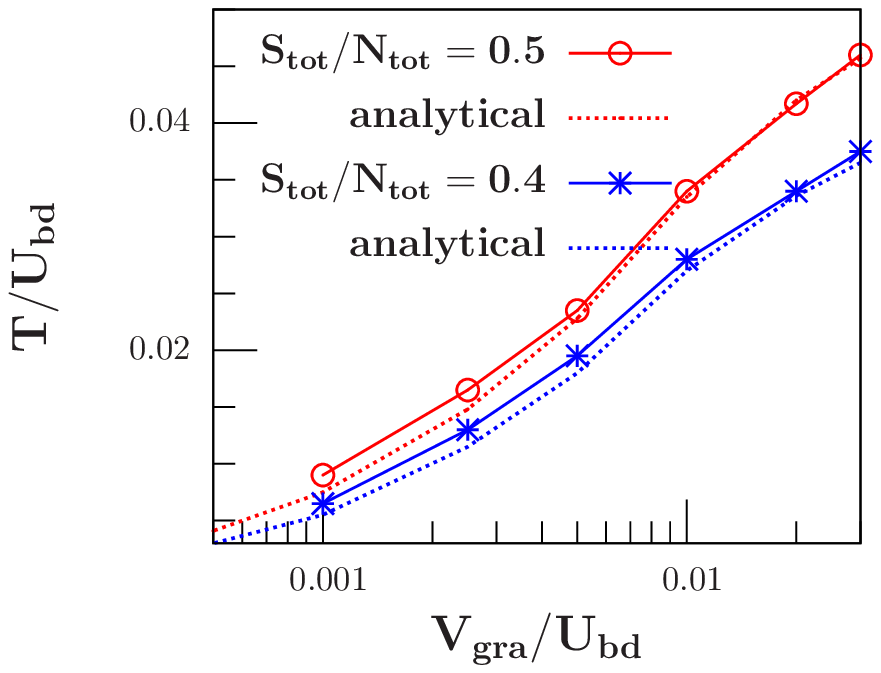}
\end{tabular}
\caption{(Color online) {\bf Upper}: Field-gradient dependence of the local entropy
per particle along the $x$ direction on the $y,z=0$ axis, for an
entropy per particle $S_{tot}/N_{tot}=0.7$. The red, green and
blue lines correspond to field gradients of $V_{gra}/U_{bd}=0.03$,
$0.01$ and $0.005$, respectively. {\bf Lower}: Adiabatic cooling
due to spin-gradient demagnetization in a cubic lattice. Data are
obtained by BDMFT+LDA and compared to the analytical
zero-tunneling approximation~\cite{D. M. Weld_2010}. Interactions
are set to $U_b=U_d=1.01U_{bd}$, and the hopping amplitudes are
$2zt_b=2zt_d=0.12U_{bd}$ for total particle number $N_{tot}
\approx 17000$ in a harmonic trap
$V_0=0.0025U_{bd}$.}\label{entropy_cooling}
\end{figure}
 The spin-gradient cooling scheme relies on the inhomogeneous entropy distribution of the system. The main effect of the demagnetization process is to decrease the local entropy per particle in the spin-mixed regions, which is essential for long-range spin order. There are three different regions corresponding to different phases of the system, namely the superfluid, spin-mixed and one-component Mott-insulating region. Initially, the superfluid and spin-mixed region carry almost all the entropy of the system, while the entropy in the one-component Mott insulator is close to zero. When the magnetic field gradient is decreased, the spin-mixed region expands, while the one-component Mott-insulating region shrinks, and the average entropy per particle in the spin mixed region is decreased. At the same time, the temperature drops, since entropy carried by hot mobile particles is drained into the expanding mixed region with a drop of local entropy per particle.
Here, we will
quantitatively establish this scenario by considering the spatial
entropy distribution and entropy-temperature relation. In the
upper panel of Fig.~\ref{entropy_cooling}, the local entropy
per particle $s/n$ is shown at different field gradient strengths for
fixed total particle number and entropy. We observe that $s/n$ decreases in the central region as the field gradient
is adiabatically decreased. Since the
number of spin excitations (with respect to the ferromagnet at zero temperature and in the presence of the field gradient) due to exchange of $\left|\uparrow \right \rangle$ and $\left|\downarrow \right \rangle$ particles between neighboring sites is increased in the demagnetization process, the total energy of the system decreases as well and, as
a result, the temperature drops from $T/U_{bd}=0.065$ to $0.035$
when the field gradient adiabatically decreases from
$V_{gra}/U_{bd}=0.03$ to $0.005$.
The resulting cooling efficiency is shown in the lower panel of
Fig. \ref{entropy_cooling}. The demagnetization cooling curve
obtained via BDMFT simulations is in good agreement with results
from the zero-tunneling limit \cite{D. M. Weld_2010}, since here
we choose the optical lattice relatively deep which makes
$t_\nu/U_{bd}$ very small. In addition, the demagnetization
cooling appears to be less efficient at larger magnetic field
gradients. This is because the strong field gradient repels
particles to the outer regions of the trap, which makes the trap
center superfluid with enhanced entropy compared to the Mott
insulator. This effect reduces the entropy capacity of the spin
degree of freedom at high field gradients.

\FloatBarrier
\section{conclusion}
In conclusion, we have investigated the thermodynamics of a
two-component Bose gas loaded into an optical lattice in the
presence of an external trap, using BDMFT+LDA and the newly
developed real-space BDMFT. We obtain the finite-temperature phase
diagram and find that at low temperature, remarkably, the system
can be {\it heated} into a Mott insulator, analogous to the
Pomeranchuk effect in $^3$He. By investigating the entropy
redistribution of the system during adiabatic spin-gradient
demagnetization, we observe efficient cooling due to entropy
transfer from the single species domains to the mixed region, and
provide a quantitative theoretical validation of recent
experiments \cite{D. M. Weld, P. Medley}. We expect our work to
provide valuable insight for realizing quantum magnetic phases in
upcoming experiments.

\begin{acknowledgments}
We acknowledge useful discussions with E. Demler, S. Kuhr, I. Titvinidze
and D. Weld. This work was supported by the China Scholarship Fund
(YL), and by the Deutsche Forschungsgemeinschaft via SFB-TR 49 and
the DIP project HO 2407/5-1. WH acknowledges the hospitality of
the Aspen Center of Physics during the final stage of this work,
supported by the National Science Foundation under Grant No.
1066293.
\end{acknowledgments}

\end{document}